\documentclass[aps,prl,reprint,superscriptaddress]{revtex4-2}
\usepackage{hyperref}
\usepackage[version =3 ]{mhchem}
\usepackage{amsmath,amssymb}
\usepackage{graphicx}
\usepackage{xcolor}

\usepackage{times}
\usepackage{fontenc}
\usepackage{epstopdf}
\usepackage{fancyhdr}
\usepackage[T1]{fontenc}
\usepackage{bm}

\begin{document}
\title{Layer Axial Phonons and Dipolar Thermal Response in Centrosymmetric Thin Films}

\author{Madhubanti Mukherjee}
\thanks{These authors contributed equally to this work.}
\affiliation{ Department of Condensed Matter and Materials Physics, S.~N. Bose National Centre for Basic Sciences, Kolkata, India }
\author{Kapil Gope}
\thanks{These authors contributed equally to this work.}
\affiliation{ Department of Condensed Matter and Materials Physics, S.~N. Bose National Centre for Basic Sciences, Kolkata, India }
\author{Barun Ghosh}
\email{bghosh@bose.res.in}
\affiliation{ Department of Condensed Matter and Materials Physics, S.~N. Bose National Centre for Basic Sciences, Kolkata, India }
\date{\today}

\begin{abstract}
In nonmagnetic centrosymmetric materials, $\mathcal{PT}$ symmetry enforces vanishing total phonon angular momentum (phonon AM) at every wave vector, making them appear unsuitable as hosts of axial phonons. Here, using first-principles calculations for BaAgAs and representative van der Waal's layered materials, we show that centrosymmetric slabs can still host large hidden layer-resolved phonon AM with both chiral and cycloidal modes. This hidden phonon AM texture produces layer resolved phonon thermal Edelstein-like responses and a real-space phonon AM dipole response, both tunable by strain. Our results establish centrosymmetric layered materials as an exciting platform for hidden axial phonon and multipolar phonon thermal response.
\end{abstract}

\maketitle

The layer index in thin-film materials is emerging as an active degree of freedom, analogous to the charge, spin, valley, and orbital indices. A common principle underlying hidden layer-related effects is that inversion-partner layers can carry large but opposite local observables. These contributions cancel when summed over the whole system. Such layer-resolved structures become observable when the layers are addressed separately or made inequivalent by a substrate, a gate potential, or a surface termination\cite{gao2021layer,hu2026half,han2025layer,dai2022quantum}. An important example is the `layer Hall effect', experimentally observed in a six-septuple-layer slab of the antiferromagnetic topological insulator MnBi$_2$Te$_4$ under an out-of-plane external electric field.~\cite{gao2021layer}. Importantly, in pristine even-layered MnBi$_2$Te$_4$, compensating layer-resolved Berry curvatures form a finite real-space dipole directly connected to the Axion magnetoelectric coupling and lead to dynamical axion quasiparticles~\cite {qiu2025observation,JA_axion}. Thus a globally compensated layer texture can have important physical significance. Layer-resolved physics has since been extended to other antiferromagnetic systems, quantum-anomalous Hall regimes, and twisted materials, giving rise to the emerging field of `layertronics'~\cite{chen2024layer,yi2024disorder,qin2026layer,anirban2023quantum,zhou2026layer,n2kv-fxf8,das2026intrinsicmagnetoelectrichalleffect}.

In parallel, lattice vibrations carrying finite phonon AM have attracted considerable research interest. The circular/elliptical motion of atoms can generate a non-zero phonon AM that can interact with the electronic spin and orbital degrees of freedom \cite{zhang2014angular,park2020phonon,PhysRevB.110.094401,PhysRevLett.128.075901,doi:10.1073/pnas.2304360121,10.1021/acsnano.4c18906,doi:10.1126/sciadv.adj4074}. These vibrational modes with a finite phonon AM have been termed chiral phonons in the past~\cite{zhang2015chiral,zhu2018observation,chen2019chiral,ishito2023truly,yokoyama2025phonon}. Recent terminology uses axial phonons as the broader class of phonon AM-carrying modes and distinguishes chiral and cycloidal motion through the helicity $h=\hat{\mathbf q}\cdot\mathbf L$ and cycloidicity $\mathbf c=\hat{\mathbf q}\times\mathbf L$, respectively~\cite{juraschek2025chiral,yang2026symmetry} (wavector ${\bf q}=q\hat{\bf q}$). Most studies of Axial phonons in nonmagnetic materials have focused on inversion-breaking systems ~\cite{suri2021chiral,grissonnanche2020chiral,moseni2022electric,pandey2018symmetry,gao2018nondegenerate,wang2022chiral,liu2017pseudospins,mishra2025chiral,pan2023vibrational} because combined $\mathcal{PT}$ symmetry ($\mathcal{P}:$ inversion, $\mathcal{T}:$ time-reversal) forces the total phonon AM of a nonmagnetic inversion-symmetric crystal to vanish at every wave vector. 

\begin{figure}[t]
\centering
\includegraphics[width=0.45\textwidth]{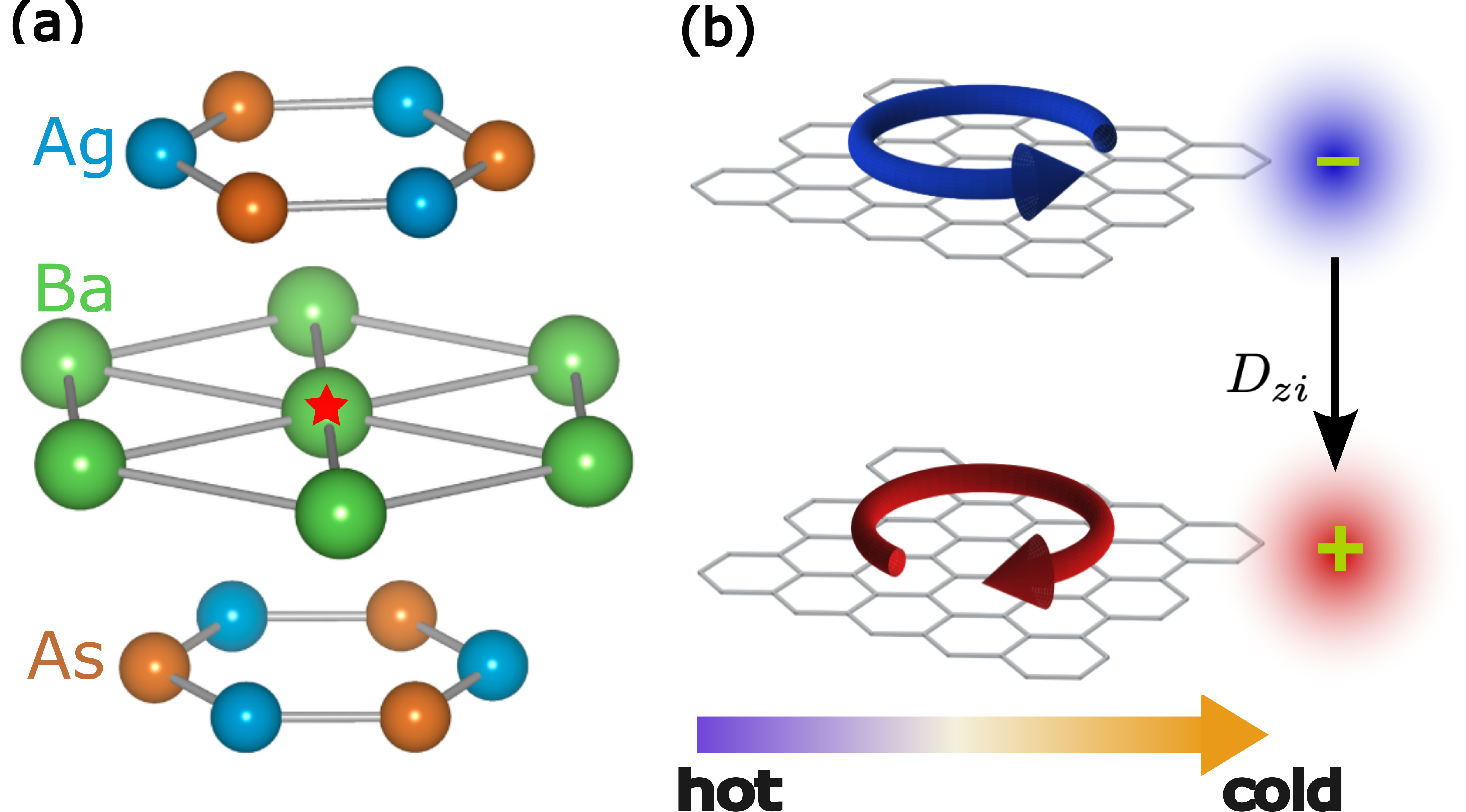}
\caption{(a) Schematic crystal structure of BaAgAs, composed of alternating Ba and Ag–As layers. The star marks the inversion center relating the top and bottom Ag–As layers. (b) Hidden layer-resolved phonon AM, illustrated by oppositely circulating arrows on the inversion-partner layers. An in-plane temperature gradient induces equal-and-opposite layer phonon AM accumulations, producing a finite real-space phonon AM dipole despite the vanishing total phonon AM.}
 \label{fig:1}
\end{figure}

\begin{figure*}[t!]
\centering
\includegraphics[width=0.95\textwidth]{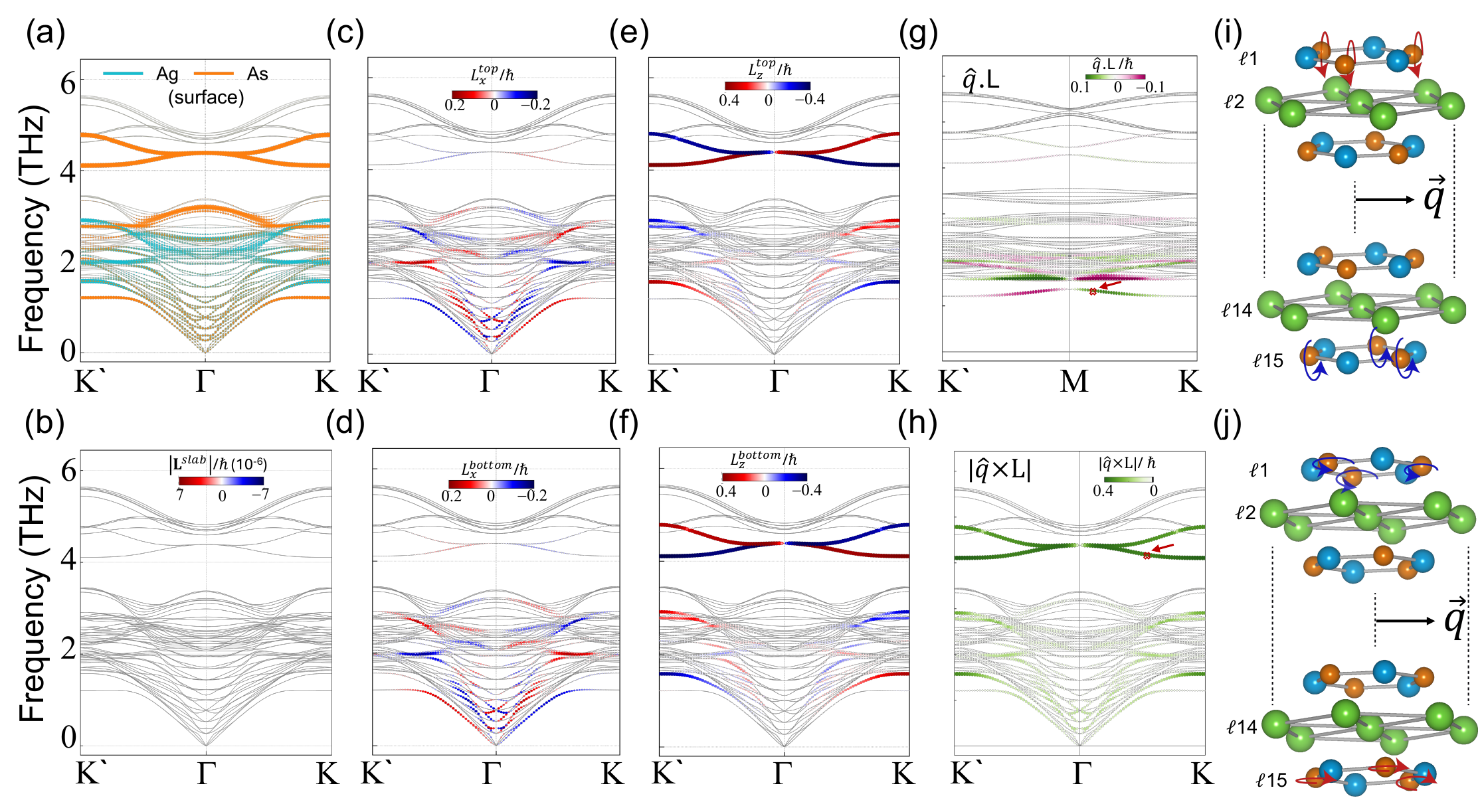}
\caption{(a) Contributions of the top- and bottom-layer Ag–As atoms to the phonon modes. (b) The total $\left|\mathbf{L}^{\mathrm{slab}}\right|$ projected onto the phonon dispersions. Clearly $\left|\mathbf{L}^{\mathrm{slab}}\right|$, and therefore each component of the phonon AM, vanishes for the whole slab. (c), (e) $L_x$ and $L_z$ components, respectively, of the phonon AM contributed by the top Ag–As layer. (d), (f) The corresponding phonon AM components for the bottom Ag–As layer. (g), (h) Helicity and cycloidicity of the top Ag--As surface, respectively. (i), (j) Representative chiral and cycloidal atomic motions, respectively, for the phonon mode (denoted by the red arrows) along $K'-M-K$ and $K'-\Gamma-K$. Red and blue circular arrows indicate opposite handedness of circular motion.}
 \label{fig:2}
\end{figure*}

A vanishing total phonon AM does not require every local contribution to vanish. Recent work has begun to uncover distinct local and compensated forms of phonon AM. Very recent examples include hidden sublattice phonon AM in ferroaxial models, sublattice-staggered phonon AM in $\mathcal{PT}$-symmetric antiferromagnets, and recent chiral surface phonons in a rocksalt structure~\cite{xie2026effective,das2026antiferro,pols2026chiral}. The layer degree of freedom as a general design principle for phonon AM in finite centrosymmetric materials remains largely unexplored. It remains unclear whether a globally centrosymmetric nonmagnetic film hosts a mode-resolved hidden phonon AM texture throughout its thickness, across distinct material and bonding classes, with both bulk derived and boundary activated contributions. Furthermore, the physical consequences of this hidden-layer phonon AM in the presence of a temperature gradient also remain unexplored.

Here, we establish hidden layer-resolved phonon AM and its associated responses in inversion-symmetric thin films using first-principles calculations. In BaAgAs slabs, inversion-partner layers carry equal and opposite mode-resolved phonon AM, while $\mathcal{PT}$ symmetry enforces a vanishing net phonon AM for the entire slab. In the periodic bulk, finite phonon AM resides only on the Ag-As layers, whereas all Ba layers carry zero phonon AM. Remarkably, in the slab geometry, not only do the Ag-As layers retain their bulk phonon AM, but all Ba layers except the inversion-centered layer also acquire finite phonon AM. Consequently, the spatial separation of these compensating contributions gives the mode-resolved phonon AM a finite, origin-independent first real-space moment. While its equilibrium Brillouin-zone sum vanishes due to $\mathcal{T}$, the non-equilibrium phonon population generated by the temperature gradient produces a layer-resolved Edelstein-like thermal phonon AM response and a nonvanishing whole-slab phonon AM dipole response. Uniaxial strain controls the allowed response components and activates otherwise forbidden ones. We obtain the same hidden layer-phonon AM texture in isostructural SrAgAs and vdW materials including four layer hBN and WSe$_2$. Notably, it also emerges in finite 1T-NiTe$_2$ slabs despite being absent in the periodic bulk and isolated monolayer, demonstrating that the effect spans distinct bonding classes and microscopic mechanisms. These results identify centrosymmetric thin films as a platform for layer-resolved axial phonons and multipolar phonon-angular-momentum responses.

The phonon AM of a mode $(\mathbf q,n)$ is given by~\cite{zhang2014angular,hamada2018phonon}, $L_{i,{\mathbf q}n}=\hbar~\bm{\epsilon}_{\mathbf qn}^{\dagger}\hat S_i\bm{\epsilon}_{\mathbf qn} =\hbar\sum_{a\in\mathrm{slab}}\bm{\epsilon}_{a,\mathbf qn}^{\dagger}\hat s_i\bm{\epsilon}_{a,\mathbf qn}$. Here, $\bm{\epsilon}_{\mathbf qn}$ is the normalized eigenvector of the complete slab, $\bm{\epsilon}_{a,\mathbf qn}$ its three Cartesian components on atom $a$, $\hat s_i$ ($i\in\{x,y,z\}$) is the spin-1 rotation generator, and $S_i$ is its block-diagonal extension over all atoms. Restricting the sum to the atoms of layer $\ell$ gives the layer resolved phonon AM as, $L_{i,{\mathbf q}n}^{\ell}=\hbar\sum_{a\in\ell}\bm{\epsilon}_{a,\mathbf qn}^{\dagger}
\hat s_i\bm{\epsilon}_{a,\mathbf qn}$. The full ${\bf L}^\mathrm{slab}$ is recovered by summing over all the layers. Inversion $\mathcal P$ exchanges layer $\ell$ with its partner $\bar\ell$ (see Fig. \ref{fig:1} (a)) and sends $\mathbf q\to-\mathbf q$; since phonon AM is an axial vector, $L_{i,{\mathbf q}n}^{\ell}=L_{i,{-\mathbf{q}}n}^{\bar\ell}$. Time reversal $\mathcal T$ also changes $\mathbf q\to-\mathbf q$ but reverses the phonon AM and leaves the layers unchanged, $L_{i,{\mathbf q}n}^{\ell}=-L_{i,{-\mathbf q}n}^{\ell}$.  Therefore, combined $\mathcal{PT}$ leads to $L_{i,{\mathbf q}n}^{\bar\ell}=-L_{i,{\mathbf q}n}^{\ell}$ and therefore, $L_{i,{\mathbf q}n}^{\mathrm{slab}}=\sum_{\ell}L_{i,{\mathbf q}n}^{\ell}=0$ for every phonon mode. Thus, although the total phonon AM vanishes in a centrosymmetric slab, inversion-partner layers can host large, equal-and-opposite phonon AM contributions. These definitions assume non-degenerate branches.
Within an exactly degenerate manifold $\mathcal{D}_g$ of degeneracy $g$, the eigenvectors are defined only up to a unitary rotation, so the layer phonon AM for an individual eigenvector is gauge-dependent. However, for each layer $\ell$, the manifold sum $L_{i,{\mathbf q}n,\mathcal{D}_g}^{\ell}=\sum_{n\in\mathcal{D}_g}L_{i,{\mathbf q}n}^\ell$ is a gauge-invariant trace. For visualization, we assign the mean $L_{i,{\mathbf q}n,\mathcal{D}_g}^{\ell}/g$ to each degenerate branch, which preserves the manifold sum.

We explicitly demonstrate this hidden phonon AM texture across different classes of materials. In the main text, we focus on BaAgAs, which adopts a layered (non-van der Waals) hexagonal structure with space group $P6_3/mmc$ (No. 194)~\cite{mardanya2019prediction,Peng2021_baagas}. The bulk unit cell contains two Ba, two Ag, and two As atoms arranged in alternating Ba triangular and Ag–As honeycomb layers [Fig ~\ref{fig:1}(a)]. The Ba layers are inversion symmetric with point group $D_{6h}$, whereas the individual Ag–As layers are noncentrosymmetric with point group $C_{3v}$. Nevertheless, the Ag–As-terminated slab is globally centrosymmetric with point group $D_{3d}$, which includes an inversion, threefold rotation $\mathcal{C}_{3z}$, and $\mathcal{M}_x$ mirror symmetry, among others. Here, the global inversion symmetry is restored by the stacking, which interchanges the Ag and As sublattices between adjacent Ag–As layers. The interlayer spacing of 2.27\r{A} between Ba and Ag-As plane is comparable (2.65 \r{A}) to the well-studied van der Waals material NiTe$_2$ [see End Matter]; therefore, the Ba and Ag-As are approximately treated as discrete layers. This crystallographic layer decomposition~\cite{Peng2021_baagas} permits a layer-resolved description of phonon angular momentum even though the crystal is three-dimensionally bonded.

The calculated phonon dispersions for the $\text{Ag-As}$-terminated slabs, comprising 15 atomic layers, are presented in Fig.~\ref{fig:2} (a-h). The absence of imaginary frequencies across the high-symmetric path confirms the dynamical stability of these slabs. The stability is further supported by {\it ab~initio} molecular dynamics calculations performed at 300K, as shown in Fig. S10. Turning to the phonon properties, the phonon spectrum exhibits a clear energy separation after 3 THz, consistent with the large mass disparity among the constituent atoms: the lighter As atoms dominate the high-frequency optical regime (4–6 THz), whereas the heavier Ag and Ba atoms dominate modes below 4 THz. Notably, up to 3 THz, pronounced hybridization occurs among the Ba, Ag, and As phonon modes. As explained earlier, combined $\mathcal{PT}$ symmetry forces the total phonon AM, $ \mathbf{L}^{\mathrm{slab}}$, to vanish throughout the Brillouin zone [Fig.~\ref{fig:2}(b)] for the entire slab. Nevertheless, the top and bottom inversion-partner layers carry large and opposite phonon AM [Figs.~\ref{fig:2}(c)–(f), Fig. S7]. The $L_z$ component shown in Fig.~\ref{fig:2}(e) dominates because of the large in-plane circular atomic motion, while $L_x$ (Fig.~\ref{fig:2}(c)) ($L_y$) remain small but finite. At the time-reversal-related $K$ and $K'$ valleys, $\mathbf{L}$ has opposite signs across the phonon branches, consistent with global $\mathcal T$ symmetry, and it is similar to the valley-contrasting nature of chiral phonons in inversion-broken transition-metal dichalcogenides.

A finite phonon AM indicates axial phonon modes, which can be further classified by the helicity $h=\hat{\mathbf q}\cdot\mathbf L$ and cycloidicity $\mathbf c=\hat{\mathbf q}\times\mathbf L$. Figs.~\ref{fig:2}(g) and (h) [see Fig.S6 for individual components of $\mathbf{c}$] show the helicity- and cycloidicity-projected phonon dispersions, respectively, for the top layer atoms of Ag–As-terminated slab. As $q_z=0$, the helicity contains only the in-plane phonon AM components and does not capture the dominant $L_z$. In contrast, cycloidicity includes $L_z$ through $c_x=q_yL_z$ and $c_y=-q_xL_z$. The modes are predominantly cycloidal along the $K'-\Gamma-K$ high-symmetry path, while mixed chiral--cycloidal character appears along the $K'-M-K$ path. The corresponding atomic motions of a predominantly chiral mode and a predominantly cycloidal mode are shown in Figs.~\ref{fig:2}(i) and (j), respectively, with the corresponding phonon wave vectors and branches indicated by red arrows in Figs.~\ref{fig:2}(g)-(h). Owing to the two-dimensional nature of the thinfilm, the predominantly chiral mode involves mainly out-of-plane atomic motion, whereas the predominantly cycloidal mode is characterized primarily by in-plane motion.

\begin{figure}
\centering
\includegraphics[width=1.0\linewidth]{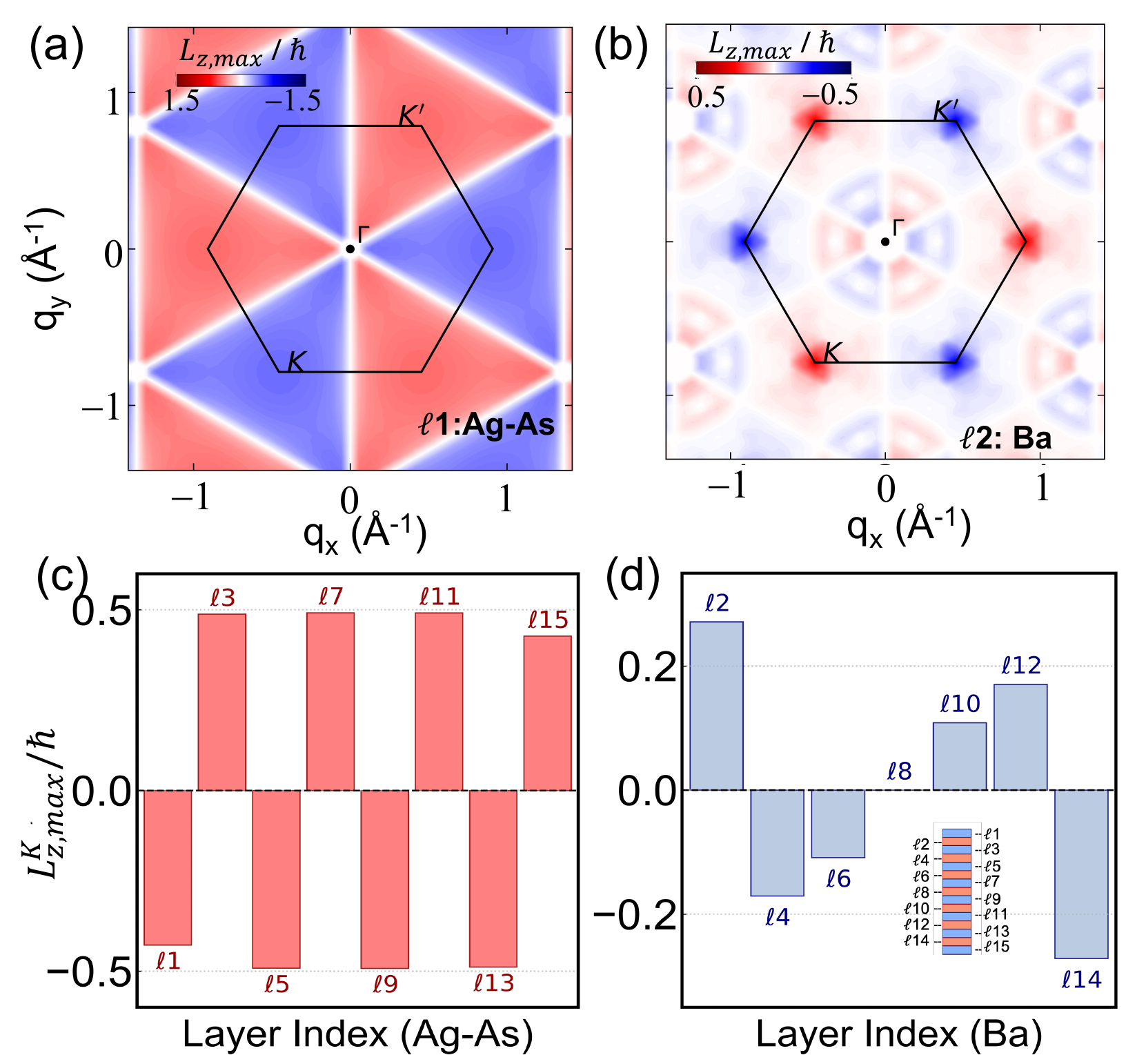}
\caption{(a) $L^\ell_z$ texture of the top Ag–As layer, $\ell_1$, branch $n=54$, $\omega=4.10$ THz (b) $L^\ell_z$ texture of the adjacent top Ba layer $\ell_2$, branch $n=28~, \omega=2.29$ THz; (c), (d) maximum $L_z$ at K versus layer index for Ag–As and Ba layers, respectively. The Ag–As layers exhibit large phonon AM throughout the slab, whereas the phonon AM in the Ba layers decays with increasing distance from the surface.}
\label{fig:3}
\end{figure}

To map the momentum-space phonon AM texture, we calculate $L_{z,{\mathbf q}n}^{\ell}$ across the two-dimensional Brillouin zone. Figs~\ref{fig:3}(a) and (b) show the phonon AM texture for the top Ag–As layers ($\ell_1$) and for the adjacent Ba layer ($\ell_2$) for branch $n=54$ with $\omega=4.10$ THz, and $n=28$ with $\omega=2.29$ THz at $K$, respectively. The maximum phonon AM of the Ag–As layers is nearly twice that of the Ba layers. This enhancement originates from the heteronuclear Ag–As network, which locally breaks in-plane inversion symmetry, unlike the monoatomic Ba triangular layers. Importantly, their corresponding inversion-partner layer carries equal but opposite phonon AM throughout the BZ. To further investigate the nature of hidden phonon AM in all the layers, we show the layer dependence of the maximum $L_z$ at the $K$ valley for the  Ag-As and Ba layers in Figures~\ref{fig:3}(c) and (d), respectively. $L_z$ remains finite in every Ag–As layer because of locally broken inversion symmetry. Importantly, for the Ba layers, $L_z$ is largest at the surfaces, decreasing toward the slab center, and vanishing at the inversion-centered Ba layer. The phonon AM of the Ba layers is strongly influenced by inversion-symmetry breaking in the local Ag--As environment. Its magnitude and decay profile are not dictated solely by symmetry; the distance from the surface also plays an important role. Thus, surface inversion breaking controls the phonon AM distribution in the Ba layers, while the intrinsic local asymmetry of the Ag--As layers sustains hidden phonon AM throughout the globally centrosymmetric slab. 

Having established the hidden phonon AM texture, we now discuss its physical consequences. We propose the layer-resolved phonon thermal Edelstein-like response, which remains finite in individual layers, and the phonon AM dipole response, which remains finite for the whole slab. The equilibrium phonon angular-momentum density is $J_i=\frac{1}{V}\sum_{\mathbf q,n}L_{i,{\bf q} n}[f_0(\omega_{{\bf q} n})+\frac{1}{2}]$, where $f_0(\omega_{{\bf q} n})=1/(e^{\hbar\omega_{{\bf q} n}/k_BT}-1)$ is the Bose distribution. At equilibrium, $\mathcal{T}$ forces the net $J_i$ to vanish. A temperature gradient, however, generates a nonequilibrium phonon population across the Brillouin zone and produces a finite layer-resolved phonon AM density. Within Boltzmann transport theory and the constant phonon relaxation-time ($\tau$) approximation, the change in the phonon distribution is $\delta f_{{\bf q} n}=-\tau v_{j,{{\bf q} n}}(\partial f_0/\partial T)(\partial T/\partial x_j)$, where $v_{j,{{\bf q} n}}=\partial\omega_n/\partial q_j$ is the $j$th component of the phonon group velocity ~\cite{hamada2018phonon}. The layer-resolved response tensor, defined through $J_i^{\ell}=\alpha_{ij}^{\ell}\partial T/\partial x_j$, is therefore,
\begin{equation}
\alpha_{ij}^{\ell}
=
-\frac{\tau}{A d_\ell N_{\bf q}}
\sum_{\mathbf q,n}
L_{i,{\bf q}n}^{\ell}
v_{j,{\bf q}n}
\frac{\partial f_0(\omega_{{\bf q} n})}{\partial T}.
\label{eq:alpha}
\end{equation}
Here, $A$ is the in-plane unit-cell area and $d_\ell$ is the average thickness of a single Ag--As or Ba layer, $N_{\bf q}$ is the number of q-points used in the BZ sum. We use an average $d_\ell=2.1~\mathring{\mathrm A}$, obtained by dividing the total slab thickness by the number of layers.

\begin{figure}[t]
\centering
\includegraphics[width=1.0\linewidth]{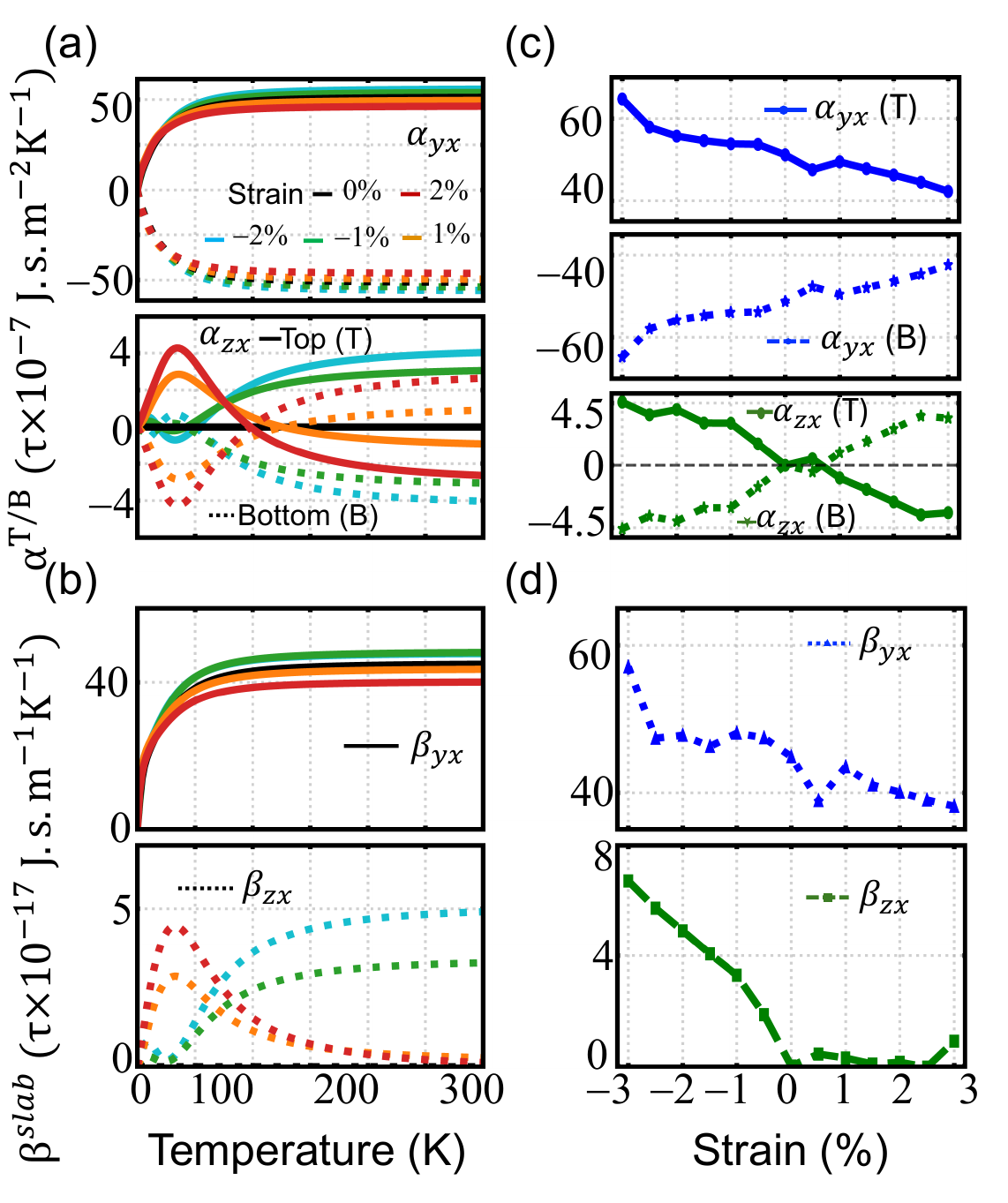}
\caption{(a) Temperature dependence of the layer-resolved $\alpha^\ell_{yx}$, $\alpha^\ell_{zx}$ for $-2\%$ to $+2\%$ strains. Solid and dotted curves denote the top and bottom layers, with $\alpha_{ij}^{\mathrm{T}}=-\alpha_{ij}^{\mathrm{B}}$. (b) Corresponding temperature dependence of slab dipole responses $\beta^{\mathrm{slab}}_{yx}$, and $\beta^{\mathrm{slab}}_{zx}$. (c, d) $\alpha_{ij}^{\ell}$ and $\beta^{\mathrm{slab}}_{ij}$ at $T=300$ K versus strain, showing the activation of symmetry-forbidden components.}
 \label{fig:4} 
\end{figure}

In the presence of $\mathcal{P}$, the $\alpha^{\mathrm{slab}}_{ij}=\sum_{\ell} \alpha_{ij}^\ell=0$, causing the response of the whole slab to vanish. Nevertheless, a layer-resolved response defined through $L_{i,{\bf q} n}^{\ell}$ can remain finite. To demonstrate this in BaAgAs, we calculate $\alpha_{ij}^{\ell}$ for the 7L-Ag-As slab terminated unstrained and under uniaxial strain ranging from $-3\%$ to $+3\%$ (Fig.~\ref{fig:4}) applied along the 100-direction of the slab. The $\mathcal{C}_{3z}$ and $\mathcal{M}_x$ symmetries of the unstrained slab supports non-zero $\alpha_{yx}^{\ell}(=-\alpha_{xy}^{\ell})$, whereas uniaxial strain breaks $\mathcal{C}_{3z}$ and activates $\alpha_{zx}^{\ell}$. Fig.~\ref{fig:4}(a) show the temperature dependence of $\alpha_{yx}^{\ell}$ and $\alpha_{zx}^{\ell}$, respectively. The $\alpha_{yx}^{\ell}$ component originates from the in-plane phonon AM components $L^\ell_y$, which is carried mainly by low-frequency phonon modes. These modes become populated at low temperatures, causing the corresponding responses to increase monotonically and saturate near $T\approx90$ K. In contrast, $\alpha_{zx}^{\ell}$ is driven by $L_z$, which extends over both low- and high-frequency modes. As higher-frequency optical modes become thermally populated, $\alpha_{zx}^{\ell}$ shows a nonmonotonic temperature dependence and saturates near $T\approx200$ K. At every temperature, the top and bottom surface layers have equal magnitudes but opposite signs, $\alpha_{ij}^{\mathrm{T}}=-\alpha_{ij}^{\mathrm{B}}$. Thus, global inversion symmetry cancels the net slab response, whereas local inversion-symmetry breaking at the surfaces produces a layer-polarized phonon AM response. Fig.~\ref{fig:4}(c) compares the tensor components as a function of strain at $T=300$ K. The $\alpha_{yx}^{\ell}$ component remains finite at zero strain, whereas $\alpha_{zx}^{\ell}$ is activated only by compressive or tensile strain. For $\tau=1$ ps, these components reach values [see SM sec IX] that may be accessible experimentally and are comparable to those reported for other chiral phonon candidate materials. \cite{wang2024chiral,pols2025chiral,yang2026symmetry, yokoyama2025phonon}. 

We next consider a response that remains finite for the complete inversion-symmetric slab under the uniform in-plane temperature gradient. Because the phonon AM on inversion-partner layers is equal and opposite but spatially separated, the slab can carry a finite real-space phonon AM dipole, defined for the $({{\bf q} n})$ mode as $D^{\mathrm{slab}}_{zi,{\bf q}n }=\sum_\ell(z_\ell-z_0)L_{i,{\bf q}n}^{\ell}$ for our slab with normal along the $z-$axis. Here, $z_\ell$ is the position of layer $\ell$ and $z_0$ is a reference plane, chosen here at the slab center. Under a shift of the reference plane: $z_0\rightarrow z_0+c$,
\begin{align}
D^{\mathrm{slab}}_{zi,{\bf q}n}(z_0+c)
&=
D^{\mathrm{slab}}_{zi,{\bf q}n}(z_0)-c\sum_{\ell}L^\ell_{i,{\bf q}n} 
=D^{\mathrm{slab}}_{zi,{\bf q}n}(z_0).
\end{align}
because $\mathcal{PT}$ symmetry enforces $\sum_{\ell}L^{\ell}_{i,{{\bf q} n}}=0$. Therefore, $D_{zi,{\bf q}n}$ is independent of the choice of origin.
The corresponding thermal-response tensor in presence of an in-plane temperature gradient is,
\begin{equation}
\beta^{\mathrm{slab}}_{ij}
=
-\frac{\tau}{AdN_{\bf q}}
\sum_{\mathbf q,n}
D_{zi,{\bf q}n}
v_{j,{\bf q}n}
\frac{\partial f_0(\omega_{{\bf q} n})}{\partial T}.
\label{eq:beta}
\end{equation}
Here, $d$ is the slab thickness, and we have dropped the z-index in $\beta$ for brevity. The $\alpha_{ij}^{\ell}$ and $\beta^{\mathrm{slab}}_{ij}$ are related by $\beta^{\mathrm{slab}}_{ij}=1/d\sum_\ell d_\ell (z_\ell-z_0) \alpha_{ij}^{\ell}$. Because $\sum_\ell\alpha_{ij}^{\ell}=0$, the response $\beta^{\mathrm{slab}}_{ij}$ is independent of the choice of origin, and remains finite for the whole slab.

Figs.~\ref{fig:4}(b) and \ref{fig:4}(d) show the temperature and strain dependence of $\beta^{\mathrm{slab}}_{ij}$, respectively. In BaAgAs slab, the symmetry-allowed components $\beta^{\mathrm{slab}}_{yx}(=-\beta^{\mathrm{slab}}_{xy})$ remain finite at zero strain, whereas $\beta^{\mathrm{slab}}_{zx}$ is activated only by compressive or tensile strain. The temperature dependence of $\beta^{\mathrm{slab}}_{ij}$ generally reflects that of the corresponding layer-resolved $\alpha_{ij}^{\ell}$, as both originate from the same nonequilibrium phonon populations, although the strain-induced components can exhibit nonmonotonic behavior. While $\alpha_{ij}^{\ell}$ depends on the composition and local environment of a particular layer, $\beta^{\mathrm{slab}}_{ij}$ is their position-weighted sum and therefore characterizes the whole slab. Uniaxial strain consequently tunes the initially allowed components of both responses and activates components forbidden in the unstrained structure.

Physically, $\alpha_{ij}^{\ell}$ describes opposite nonequilibrium phonon AM accumulations on inversion-partner layers, whereas $\beta^{\mathrm{slab}}_{ij}$ measures their spatial dipole. Similar to how $\alpha^\ell_{ij}$ can be related to the phonon magnetization induced by a temperature gradient, the components of $\beta^{\mathrm{slab}}_{ij}$ can be related to the multipolar magnetic moments including the toroidal and quadrupolar moments. Moreover, similar to the rigid-body rotation associated with a net thermal phonon Edelstein-like response observed recently in chiral Te~\cite{zhang2025measurement}, local (within layer) relaxation of the compensated layer phonon AM could in-principle generate opposite torques, driving internal torsional or bending modes without net rotation. Possible signatures could therefore include opposite surface magneto-optical responses, interfacial spin or orbital accumulation, and torque-dipole driven mechanical motion ~\cite{kim2023chiral,ma2024chiral,nabei2026orbital,fransson2023chiral,sato2026orbital,xzgb-33w5}. In metallic BaAgAs [Fig. S11], electron--phonon coupling could partially transfer lattice angular momentum to electronic spin or orbital degrees of freedom, potentially providing a possible electrical detection channel, although electronic thermoelectric backgrounds would need to be distinguished.

We establish the layer index as an active degree of freedom for phonon AM in globally inversion-symmetric systems. In BaAgAs, inversion-partner layers carry equal and opposite phonon AM, giving $\mathbf L^{\mathrm{slab}}=0$, while each layer except the inversion-centered layer retains finite phonon AM. The compensated texture produces opposite layer-resolved thermal phonon AM responses and a nonvanishing real-space phonon AM dipole response for the whole slab. Uniaxial strain further tunes the allowed response components and activates otherwise forbidden components. The same phenomenon is observed for different surface terminations, different materials including iso-structural SrAgAs and in vdW-materials including four-layer hBN and WSe$_2$ [see sec S1-S5] proving the validity of layer Axial phonons across different bonding classes.  Our results also broaden the materials scope to include inversion-symmetric thin films as exciting material candidates for hidden layer phonon AM and its associated responses.

\section{Acknowledgements}
We thank Swati Chaudhary and Arijit Haldar for helpful discussions. B. G. acknowledges the SEED Grant provided by SNBNCBS and Prime Minister Early Career Research Grant (PM-ECRG) from Anusandhan National Research Foundation (ANRF), file number ANRF/ECRG/2024/003677/PMS. We acknowledge the National Supercomputing Mission (NSM) for providing computing resources of ‘PARAM RUDRA’ at SNBNCBS, Salt Lake, Kolkata-700106, India. 

\bibliography{manuscript}

\onecolumngrid
\begin{center}
\textbf{End Matter}
\end{center}
\twocolumngrid
\appendix

\begin{figure}[h]
\centering
\includegraphics[width=1.0\linewidth]{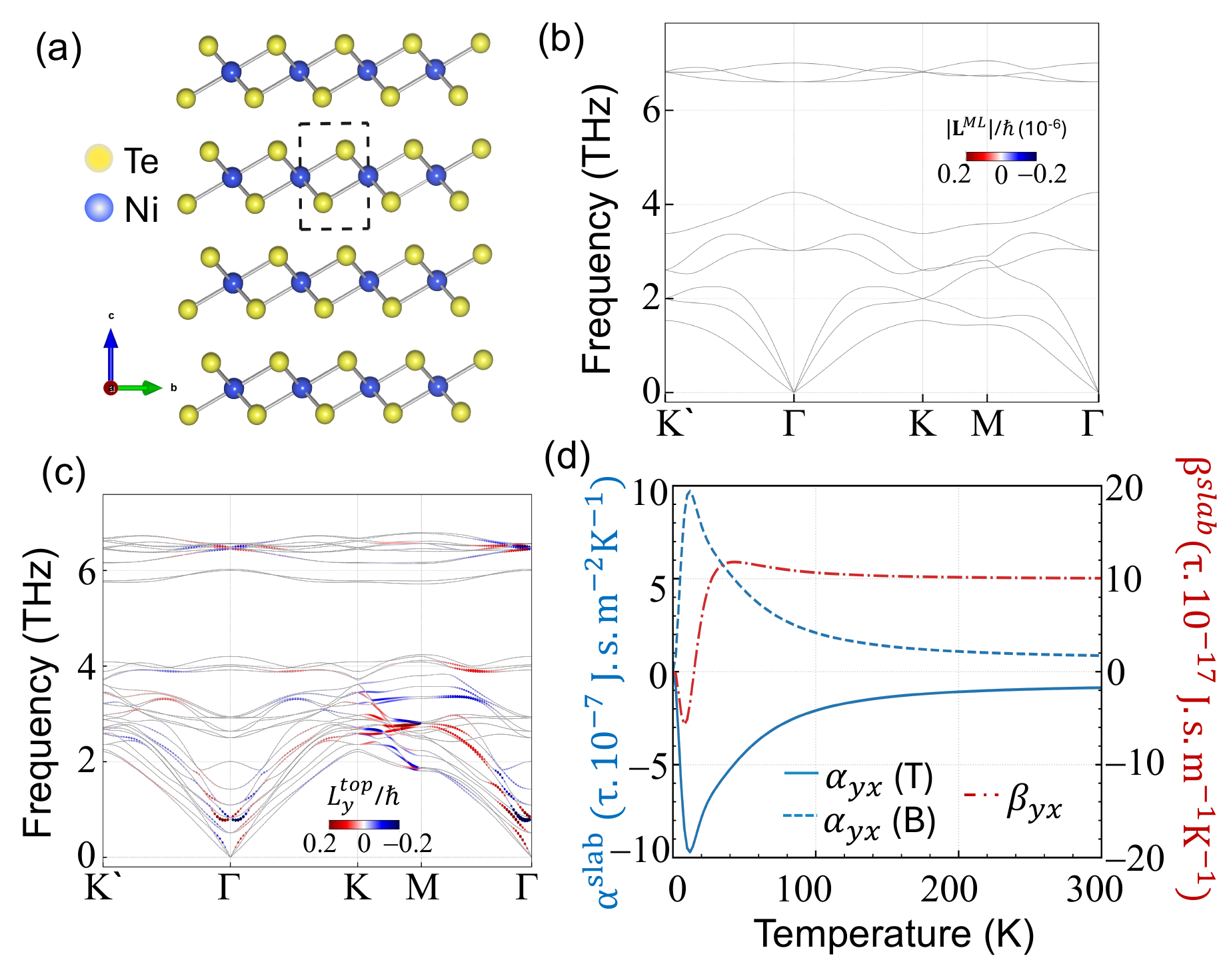}
\caption{(a) Side view of a four-layer NiTe$_2$ slab with $P\bar{3}m1$ symmetry; the dashed box marks one monolayer. (b) Total phonon AM magnitude projected onto the monolayer phonon dispersion, which vanishes within numerical precision. (c) The $y$ component of the top-layer phonon AM, $L_y^{\mathrm{top}}$, projected onto the phonon dispersion of the fourlayer slab, showing finite hidden layer phonon AM. (d) Temperature dependence of the layer-resolved responses $\alpha^\ell_{yx}$ for the top (T) and bottom (B) layers and the whole-slab dipole response $\beta^{\mathrm{slab}}_{yx}$.
\label{fig:5}}
\end{figure}

So far, we have considered the non-vdW compound BaAgAs, where the Ag-As and the Ba planes are treated as distinct layers. In van-der-Waals materials such as NiTe$_2$, h-BN, and WSe$_2$ (see SM for h-BN and WSe$_2$), the individual building-block layers are well separated by the vdW gap, providing a distinct definition of a layer unit. A NiTe$_2$ monolayer thus denotes one Te-Ni-Te unit. Both the bulk crystal and the isolated monolayer of 1T-NiTe$_2$ belong to the same centrosymmetric $P\bar{3}m1$ structure (space group no.~164)~\cite{ghosh2019observation,hlevyack2021dimensional,nappini2020transition}. Each monolayer comprises a Ni atom sandwiched between two Te atoms, as indicated by the dashed box in Fig.~\ref{fig:5}(a). In both the monolayer and the bulk, the two Te atoms are inversion partners, with Ni located at the inversion center. Combined $\mathcal{PT}$ symmetry therefore forces the total phonon AM of both the periodic bulk and the isolated monolayer to vanish at every $\mathbf q$, as shown for the monolayer in Fig.~\ref{fig:5}(b).

Remarkably, when four such monolayers are assembled into a finite slab, a given layer is no longer an inversion-invariant subsystem. In this case, the spatial inversion exchanges a layer with a distinct partner layer on the opposite side of the slab. The inequivalent out-of-plane environments produced by the slab geometry can therefore activate finite layer phonon AM, as illustrated by $L_y^{\mathrm{top}}$ in Fig.~\ref{fig:5}(c). Combined $\mathcal{PT}$ symmetry makes the contributions from inversion-partner layers equal and opposite, such that the total slab phonon AM remains zero, $\mathbf L^{\mathrm{slab}}=0$. In addition to the $L_y$ component shown here, the slab also supports finite $L_x$ and $L_z$ components. Consequently, the four layer NiTe$_2$ slab exhibits finite layer-resolved phonon thermal Edelstein-like responses $\alpha_{ij}^{\ell}$ and a nonzero whole-slab real-space dipole response $\beta_{ij}^{\mathrm{slab}}$ [Fig.~\ref{fig:5}(d)]. Thus, dimensional reduction from the periodic bulk to a finite multilayer slab can activate hidden layer phonon AM and its associated multipolar thermal response even when both the bulk and isolated monolayer host vanishing phonon AM. This is distinct from other vdW materials such as h-BN and WSe$_2$, where the individual monolayer units themselves possess finite PAM because of their in-plane inversion asymmetry.

\begin{figure}[b]
\centering
\includegraphics[width=0.9\linewidth]{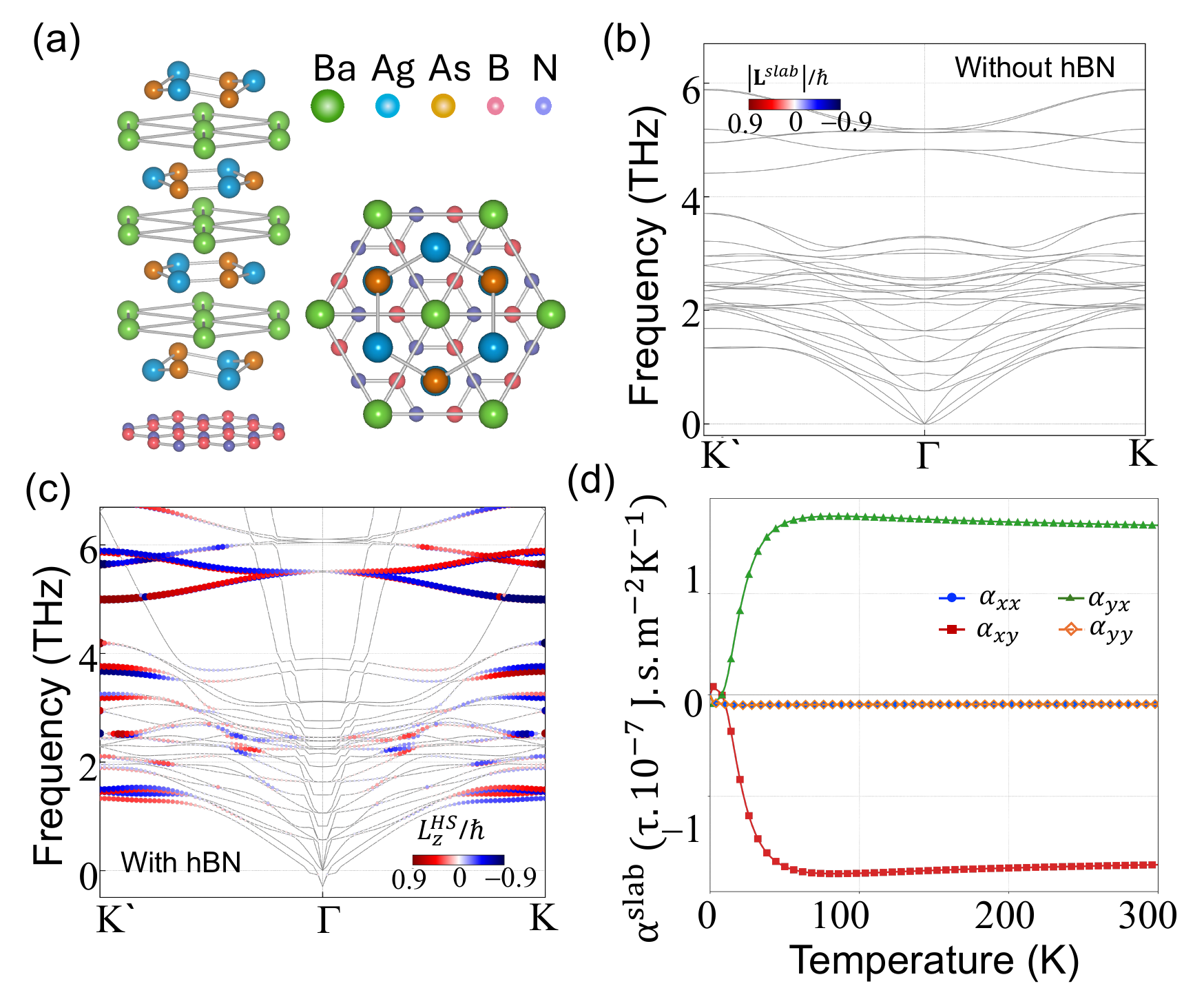}
\caption{(a) side and top views of the $\text{hBN/BaAgAs}$ heterostructure formed by placing a symmetric 7 layers $\text{BaAgAs}$ slab on a monolayer $\text{hBN}$ substrate, (b) total phonon AM ($L_z$) projected onto the phonon dispersion of the symmetric bare BaAgAs slab without h-BN, where inversion is not broken, (c) z-component of total phonon AM ($L_z$) projected onto the phonon dispersion of the heterostructure, highlighting finite angular momentum across the whole slab, and (d) the calculated thermal response $\alpha_{ij}$ in the heterostructure slab due to net nonzero ${\bf L}^{\mathrm{slab}}$.}
 \label{fig:6}
\end{figure}

Finally, for completeness, we discuss a practical route of revealing a hidden phonon AM in an otherwise centrosymmetric slab. Taking the example of BaAgAs, we interface it with a monolayer hexagonal boron nitride (hBN) substrate [Fig. ~\ref{fig:6}(a)]. We emphasize that this is not essential for obtaining the responses $\alpha_{ij}^{l}$ and $\beta^{\mathrm{slab}}_{ij}$ discussed in the preceding section. Owing to the small lattice mismatch ($\sim 3.8\%$), the $\sqrt{3}\times\sqrt{3}$ hBN/BaAgAs heterostructure forms a stable interface free of soft or imaginary vibrational modes [Fig. ~\ref{fig:6}(c)]. As expected, phonon dispersion has a large mass difference among the constituent elements. The lighter B and N atoms dominate the high-frequency optical branches ($>10$~THz), while the low-energy modes remain governed by Ag and As vibrations.

As expected, the bare symmetric BaAgAs slab carries zero total phonon AM [Fig. ~\ref{fig:6}(b)]. The substrate-induced inversion-symmetry breaking, by contrast, removes the constraint enforcing the layer cancellation and generates a strong net ${\bf L}^{\mathrm{slab}} \neq 0$, with finite L$_x$, L$_y$, and L$_z$ in the presence of hBN. Fig. ~\ref{fig:6}(c) shows the z-component of the total phonon AM; however, due to finite thickness, the slab has considerable L$_x$ and L$_y$. 

A nonzero response follows from the finite ${\bf L}^{\mathrm{slab}}$. The heterostructure belongs to the chiral point group $3$ (space group No.~143), which lacks mirror, inversion, and rotoinversion symmetries; the allowed components are consequently $\alpha_{xy}(=-\alpha_{yx})$ together with the diagonal $\alpha_{xx}(=\alpha_{yy})$. This is confirmed by our \emph{ab initio} response calculations in Fig. ~\ref{fig:6}(d). Similar to $\alpha_{ij}^{\ell}$, the components $\alpha_{ij}$ increase with temperature at low $T$ as the Axial phonon population builds up, then saturate at high T. 

\end{document}